\documentclass{article}%
\usepackage{amsfonts}
\usepackage{amsmath}
\usepackage{amssymb}
\usepackage{graphicx}%
\providecommand{\U}[1]{\protect\rule{.1in}{.1in}}

\begin{document}

\title{Chern-Simons Supergravity in D=3 and Maxwell superalgebra}
\author{P. K. Concha$^{1,2,3}$, O. Fierro$^{4}$, E. K. Rodr\'{\i}guez$^{1,2,3}$, P.
Salgado$^{1}$\\$^{1}${\small {\textit{Departamento de F\'{\i}sica, Universidad de
Concepci\'{o}n,}} }\\{\small {\textit{Casilla 160-C, Concepci\'{o}n, Chile.}} }\\\ $^{2}${\small {\textit{Dipartimento di Scienza Applicata e Tecnologia
(DISAT),}} }\\{\small {\textit{Politecnico di Torino, Corso Duca degli Abruzzi 24,}}}\\{\small {\textit{I-10129 Torino, Italia.}} }\\$^{3}${\small {\textit{Instituto Nazionale di Fisica Nucleare (INFN) Sezione
di Torino,}} }\\{\small {\textit{Via Pietro Giuria 1, 10125 Torino, Italia.}}}\\$^{4}${\small {\textit{Departamento de Ciencias Fisicas, Universidad Andres
Bello,}} }\\{\small {\textit{Republica 220, Santiago, Chile.}} }\\
\\{\small {\textit{E-mail:} {\texttt{patrickconcha@udec.cl},
\texttt{o.fierro@uandresbello.edu}}},}\\{\small {{ \texttt{everodriguez@udec.cl}, \texttt{pasalgad@udec.cl}}}}}
\maketitle

\begin{abstract}
We present the construction of the $D=3$ Chern-Simons supergravity action
without cosmological constant from the minimal Maxwell superalgebra
$s\mathcal{M}_{3}$. \ This superalgebra contains two Majorana fermionic
charges and can be obtained from the $\mathfrak{osp}\left(  2|1\right)
\otimes\mathfrak{sp}\left(  2\right)  $ superalgebra using the abelian
semigroup expansion procedure. \ The components of the Maxwell invariant
tensor are explicitly derived.

\end{abstract}

\section{Introduction}

\qquad There is a particular interest in (super)gravity theories in modifying
the Poincar\'{e} symmetries to bigger ones. \ A well known enlargement of the
Poincar\'{e} algebra is the Maxwell algebra $\mathcal{M}$ which is obtained by
adding a constant electromagnetic field background to the Minkowski spacetime
\cite{BCR, Schrader}. \ This algebra is caracterized by the introduction of
tensorial generators $Z_{ab}$ which modify the commutation relation of the
translation generators $P_{a}$,%
\begin{equation}
\left[  P_{a},P_{b}\right]  =Z_{ab}.
\end{equation}
Based on the $D=4$ Maxwell symmetries, it was recently shown an alternative
way to introduce a generalized cosmological term to a gravity action
\cite{AKL}. \ Furthermore, it was recently pointed out that the Maxwell type
algebras allow to recover General Relativity from Chern-Simons (CS) and
Born-Infeld (BI) gravity theories \cite{GRCS, CPRS1, CPRS2, CPRS3}.

Interestingly, the supersymmetric extension of the Maxwell algebra describes
the geometry of a four dimensional $\mathcal{N}=1$ superspace in the presence
of a constant abelian supersymmetric gauge field background \cite{BGKL}.
\ This modifies the superMinkowski space into the superMaxwell space. \ In
particular, the $D=4$ minimal Maxwell superalgebra introduced in \cite{BGKL}
contains the usual Maxwell algebra as subalgebra. \ This minimal Maxwell
superalgebra and its generalizations have been extensively studied using the
expansion methods in refs. \cite{AILW, CR1}. \ These superalgebras have the
particularity to have more than one spinor charge and can be viewed as the
generalization of the D'Auria-Fr\'{e} superalgebra and the Green algebras
introduced respectively in refs. \cite{AF, Green}.

In particular, the minimal Maxwell superalgebra is obtained after a resonant
$S$-expansion of $\mathfrak{osp}\left(  4|1\right)  $ superalgebra \cite{CR1}.
Subsequently, as shown in ref. \cite{CR2}, pure supergravity can be derived as
a MacDowell-Mansouri like action from the Maxwell symmetries. \

The $S$-expansion method is a powerful tool in order to derive new Lie
(super)algebras and build new (super)gravity theories. \ Basically, it
consists in combining the structure constants of a Lie (super)algebra
$\mathfrak{g}$ with the multiplication law of a semigroup $S$ \cite{Sexp}. \ A
very useful advantage of this procedure is that it provides with an invariant
tensor for the $S$-expanded (super)algebra $\mathfrak{G}=S\times\mathfrak{g}$
in terms of an invariant tensor for the original (super)algebra $\mathfrak{g}%
$. \ In particular, the invariant tensor is a crucial ingredient in the
construction of (super)gravity actions. \ Some interesting applications of the
$S$-expansion method in (super)gravity theories can be found in refs.
\cite{GRCS, CPRS1, CPRS2, CPRS3, CR2, IRS1, GSRS, Topgrav}.

An interesting formalism which allows to construct a gauge theory of
supergravity in odd dimensions is the Chern-Simons approach. \ A CS gravity
theory has the advantage to be a "gauge" theory of gravity whose spin
connection and vielbein can be seen as independent fields.\ \ In particular, a
good candidate to describe a three-dimensional CS supergravity theory with a
cosmological constant is the $AdS$ supergroup. \ The most generalized
supersymmetric extension of the three-dimensional $AdS$ algebra is given by
the direct product \cite{AchuTown}
\begin{equation}
\mathfrak{osp}\left(  2|p\right)  \otimes\mathfrak{osp}\left(  2|q\right)  ,
\end{equation}
describing a $\left(  p,q\right)  $-type $AdS$-Chern-Simons supergravity in
presence of a cosmological constant. \ The CS supergravity action is
constructed out of the connection one-form $A$ associated to the $AdS$
supergroup as follows \cite{Cham1, Cham2}%
\begin{equation}
S=k\int\left\langle A\left(  dA+\frac{2}{3}A^{2}\right)  \right\rangle ,
\end{equation}
where $\left\langle \cdots\right\rangle $ denotes the invariant tensor. \

Interestingly, the $\mathfrak{osp}\left(  2|p\right)  \otimes\mathfrak{osp}%
\left(  2|q\right)  $ superalgebra allows to construct a non minimal
three-dimensional $AdS$ CS supergravity theory. \ In particular, the minimal
$AdS$ CS supergravity is obtained when $p=1$ and $q=0$ $\left(  \mathfrak{osp}%
\left(  2|1\right)  \otimes\mathfrak{sp}\left(  2\right)  \right)  $
\cite{GTW}. \ As was pointed out in ref. \cite{AchuTown}, the presence of
$\mathcal{N}=p+q$ supersymmetries allows to introduce CS terms related to the
$O\left(  p\right)  \otimes O\left(  q\right)  $ gauge symmetry. \

The purpose of this paper is to construct a $D=3$ minimal Maxwell-Chern-Simons
supergravity action. \ To this aim, the $S$-expansion of the $\mathfrak{osp}%
\left(  2|1\right)  \otimes\mathfrak{sp}\left(  2\right)  $ superalgebra is
considered in order to derive the minimal Maxwell superalgebra $s\mathcal{M}%
_{3}$. \ The procedure considered here allows to find the non-vanishing
components of a Maxwell invariant tensor which are required to construct a CS
action. \ The CS formalism used here represents a toy model in order to
approach problems present in higher dimensions or in higher $\mathcal{N}%
$-extended supergravity theories.

This work is organized as follows: in section 2 we briefly review the CS
supergravity theory for the $\mathfrak{osp}\left(  2|1\right)  \otimes
\mathfrak{sp}\left(  2\right)  $ superalgebra. \ Section 3 contains our main
results. \ We obtain the minimal Maxwell superalgebra $s\mathcal{M}_{3}$ from
the $\mathfrak{osp}\left(  2|1\right)  \otimes\mathfrak{sp}\left(  2\right)  $
superalgebra using the $S$-expansion procedure. \ The components of a
superMaxwell invariant tensor are presented and a CS supergravity action is
constructed. \ Section 4 concludes the work with some comments about the
Maxwell supersymmetries and possible developments.

\section{$D=3$ Chern-Simons supergravity theory and $AdS$ superalgebra}

In this section we briefly review the construction of the Chern-Simons
supergravity action in $D=3$ for the $AdS$ superalgebra, $\mathfrak{osp}%
\left(  2|1\right)  \mathfrak{\otimes sp}\left(  2\right)  $. \ The
(anti)commutation relations for this superalgebra are given by
\begin{align}
\left[  \tilde{J}_{ab},\tilde{J}_{cd}\right]   &  =\eta_{bc}\tilde{J}%
_{ad}-\eta_{ac}\tilde{J}_{bd}-\eta_{bd}\tilde{J}_{ac}+\eta_{ad}\tilde{J}%
_{bc},\label{ads1}\\
\left[  \tilde{J}_{ab},\tilde{P}_{c}\right]   &  =\eta_{bc}\tilde{P}_{a}%
-\eta_{ac}\tilde{P}_{b},\text{ \ \ \ \ }\\
\text{\ \ \ }\left[  \tilde{P}_{a},\tilde{P}_{b}\right]   &  =\tilde{J}%
_{ab},\\
\left[  \tilde{P}_{a},\tilde{Q}_{\alpha}\right]   &  =-\frac{1}{2}\left(
\Gamma_{a}\tilde{Q}\right)  _{\alpha},\\
\left[  \tilde{J}_{ab},\tilde{Q}_{\alpha}\right]   &  =-\frac{1}{2}\left(
\Gamma_{ab}\tilde{Q}\right)  _{\alpha},\\
\left\{  \tilde{Q}_{\alpha},\tilde{Q}_{\beta}\right\}   &  =-\frac{1}%
{2}\left[  \left(  \Gamma^{ab}C\right)  _{\alpha\beta}\tilde{J}_{ab}-2\left(
\Gamma^{a}C\right)  _{\alpha\beta}\tilde{P}_{a}\right]  , \label{ads6}%
\end{align}
where $\tilde{J}_{ab}$, $\tilde{P}_{a}$ and $\tilde{Q}_{\alpha}$ are the
generators of Lorentz transformations, the $AdS$ boost and supersymmetry,
respectively. Here $C$ stands for the charge conjugation matrix, $\Gamma_{a}$
are Dirac matrices and $\Gamma_{ab}=\frac{1}{2}\left[  \Gamma_{a},\Gamma
_{b}\right]  $.

The Chern-Simons action in $\left(  2+1\right)  $ dimensions \cite{Cham1,
Cham2} is given by
\begin{equation}
S_{CS}^{\left(  2+1\right)  }=k\int\left\langle A\left(  dA+\frac{2}{3}%
A^{2}\right)  \right\rangle . \label{actm}%
\end{equation}
Here, $A$ corresponds the one-form gauge connection for the $\mathfrak{osp}%
\left(  2|1\right)  \mathfrak{\otimes sp}\left(  2\right)  $ superalgebra%
\begin{equation}
A=\frac{1}{2}\omega^{ab}\tilde{J}_{ab}+\frac{1}{l}e^{a}\tilde{P}_{a}+\frac
{1}{\sqrt{l}}\psi^{\alpha}\tilde{Q}_{\alpha}, \label{oneform}%
\end{equation}
whose associated curvature two-form $F=dA+A\wedge A$ is
\begin{equation}
F=F^{A}T_{A}=\frac{1}{2}\mathcal{R}^{ab}\tilde{J}_{ab}+\frac{1}{l}R^{a}%
\tilde{P}_{a}+\frac{1}{\sqrt{l}}\Psi^{\alpha}\tilde{Q}_{\alpha},
\end{equation}
where%
\begin{align*}
\mathcal{R}^{ab}  &  =d\omega^{ab}+\omega_{\text{ }c}^{a}\omega^{cb}+\frac
{1}{l^{2}}e^{a}e^{b}+\frac{1}{2l}\bar{\psi}\Gamma^{ab}\psi,\\
R^{a}  &  =de^{a}+\omega_{\text{ }b}^{a}e^{b}-\frac{1}{2}\bar{\psi}\Gamma
^{a}\psi,\\
\Psi &  =\nabla\psi=d\psi+\frac{1}{4}\omega_{ab}\Gamma^{ab}\psi+\frac{1}%
{2l}e^{a}\Gamma_{a}\psi.
\end{align*}
In $\left(  \ref{actm}\right)  $ the bracket $\left\langle \cdots\right\rangle
$ stands for the non-vanishing components of an invariant tensor for the
$\mathfrak{osp}\left(  2|1\right)  \mathfrak{\otimes sp}\left(  2\right)  $
superalgebra in $\left(  2+1\right)  $-dimensions:%
\begin{align}
\left\langle \tilde{J}_{ab}\tilde{J}_{cd}\right\rangle  &  =\mu_{0}\left(
\eta_{ad}\eta_{bc}-\eta_{ac}\eta_{bd}\right)  ,\label{Inv1}\\
\left\langle \tilde{J}_{ab}\tilde{P}_{c}\right\rangle  &  =\mu_{1}%
\epsilon_{abc},\\
\left\langle \tilde{P}_{a}\tilde{P}_{b}\right\rangle  &  =\mu_{0}\eta_{ab},\\
\left\langle \tilde{Q}_{\alpha}\tilde{Q}_{\beta}\right\rangle  &  =\left(
\mu_{0}-\mu_{1}\right)  C_{\alpha\beta}, \label{Inv4}%
\end{align}
\ where $\mu_{0}$ and $\mu_{1}$ are arbitrary constants.

Considering $\left(  \ref{Inv1}\right)  $-$\left(  \ref{Inv4}\right)  $ and
the one-form connection $\left(  \ref{oneform}\right)  $, the CS action
$\left(  \ref{actm}\right)  $ for the $\mathfrak{osp}\left(  2|1\right)
\mathfrak{\otimes sp}\left(  2\right)  $ superalgebra can be written as
\begin{align}
S_{CS}^{\left(  2+1\right)  }  &  =k\int_{M}\frac{\mu_{0}}{2}\left(
\omega_{\text{ }b}^{a}d\omega_{\text{ }a}^{b}+\frac{2}{3}\omega_{\text{ }%
c}^{a}\omega_{\text{ }b}^{c}\omega_{\text{ }a}^{b}+\frac{2}{l^{2}}e^{a}%
T_{a}+\frac{2}{l}\bar{\psi}\Psi\right) \nonumber\\
&  +\frac{\mu_{1}}{l}\left(  \epsilon_{abc}\left(  R^{ab}e^{c}+\frac{1}%
{3l^{2}}e^{a}e^{b}e^{c}\right)  -\bar{\psi}\Psi\right)  -d\left(  \frac
{\mu_{1}}{2l}\epsilon_{abc}\omega^{ab}e^{c}\right)
\end{align}
where $T^{a}=de^{a}+\omega_{\text{ }b}^{a}e^{b}$ is the torsion 2-form. \ This
action describes the most general $\mathcal{N}=1,$ $D=3$ Chern-Simons
supergravity action (with cosmological constant) for the $AdS$ supergroup
\cite{GTW}.

\section{The minimal Maxwell superalgebra and CS Supergravity action}

\qquad The abelian semigroup expansion method ($S$-expansion) is a powerful
tool in order to obtain new Lie (super)algebras from known ones \cite{Sexp,
AMNT}. \ The $S$-expansion procedure consists in combining the multiplication
law of a semigroup $S$ with the structure constants of a Lie algebra
$\mathfrak{g}$. \ The new Lie algebra $\mathfrak{G}=S\times\mathfrak{g}$ is
called the $S$-expanded algebra.

In this section, we show that a three-dimensional minimal Maxwell superalgebra
$s\mathcal{M}_{3}$ can be derived from the $\mathfrak{osp}\left(  2|1\right)
\mathfrak{\otimes sp}\left(  2\right)  $ superalgebra using the $S$-expansion
procedure with a particular choice of a semigroup $S$. \ The result obtained
here will be useful in the construction of a Maxwell-Chern-Simons supergravity
in $D=3$ which we shall approach in the next subsection.

A necessary step before applying the $S$-expansion method consists in
considering a decomposition of the original algebra $\mathfrak{g}%
=\mathfrak{osp}\left(  2|1\right)  \otimes\mathfrak{sp}\left(  2\right)  $ in
subspaces $V_{p}$,%
\begin{equation}
\mathfrak{g=osp}\left(  2|1\right)  \mathfrak{\otimes sp}\left(  2\right)
=V_{0}\oplus V_{1}\oplus V_{2}%
\end{equation}
where $V_{0}$ corresponds to the Lorentz subalgebra which is generated by
$\tilde{J}_{ab}$, $V_{1}$ corresponds to the supersymmetry translation
generated by $\tilde{Q}_{\alpha}$ and $V_{2}$ corresponds to the AdS boost
generated by $\tilde{P}_{a}$. \ The subspace structure may be written as%
\begin{equation}%
\begin{tabular}
[c]{ll}%
$\left[  V_{0},V_{0}\right]  \subset V_{0},$ & $\left[  V_{1},V_{1}\right]
\subset V_{0}\oplus V_{2},$\\
$\left[  V_{0},V_{1}\right]  \subset V_{1},$ & $\left[  V_{1},V_{2}\right]
\subset V_{1,}$\\
$\left[  V_{0},V_{2}\right]  \subset V_{2},$ & $\left[  V_{2},V_{2}\right]
\subset V_{0}.$%
\end{tabular}
\ \ \ \ \ \label{sdes}%
\end{equation}

The next step consists in finding a subset decomposition of a semigroup $S$
which is "resonant" with respect to the subspace structure $\left(
\ref{sdes}\right)  $. \ Let us consider $S_{E}^{(4)}=\left\{  \lambda
_{0},\lambda_{1},\lambda_{2},\lambda_{3},\lambda_{4},\lambda_{5}\right\}  $ as
the relevant finite abelian semigroup whose elements are dimensionless and
obey the multiplication law%
\begin{equation}
\lambda_{\alpha}\lambda_{\beta}=\left\{
\begin{array}
[c]{c}%
\lambda_{\alpha+\beta}\text{, \ \ \ \ when }\alpha+\beta\leq5,\\
\lambda_{5}\text{, \ \ \ \ \ \ \ when }\alpha+\beta>5.
\end{array}
\right.  \label{ls1}%
\end{equation}
Here $\lambda_{5}$ plays the role of the zero element of the semigroup
$S_{E}^{\left(  4\right)  }$, so we have for each $\lambda_{\alpha}\in
S_{E}^{\left(  4\right)  },$ $\lambda_{5}\lambda_{\alpha}=\lambda_{5}=0_{s}$.
\ Let us consider a subset decomposition $S_{E}^{(4)}=S_{0}\cup S_{1}\cup
S_{2},$ with%
\begin{align}
S_{0}  &  =\left\{  \lambda_{0},\lambda_{2},\lambda_{4},\lambda_{5}\right\}
,\\
S_{1}  &  =\left\{  \lambda_{1},\lambda_{3},\lambda_{5}\right\}  ,\\
S_{2}  &  =\left\{  \lambda_{2},\lambda_{4},\lambda_{5}\right\}  .
\end{align}
This subset decomposition is said to be "resonant" since it satisfies the same
structure as the subspaces $V_{p}$ [compare with eqs. $\left(  \ref{sdes}%
\right)  $]%
\begin{equation}%
\begin{tabular}
[c]{ll}%
$S_{0}\cdot S_{0}\subset S_{0},$ & $S_{1}\cdot S_{1}\subset S_{0}\cap S_{2}%
,$\\
$S_{0}\cdot S_{1}\subset S_{1},$ & $S_{1}\cdot S_{2}\subset S_{1},$\\
$S_{0}\cdot S_{2}\subset S_{2},$ & $S_{2}\cdot S_{2}\subset S_{0}.$%
\end{tabular}
\end{equation}
Following theorem IV.2 of ref. \cite{Sexp}, we can say that the superalgebra
\begin{equation}
\mathfrak{G}_{R}=W_{0}\oplus W_{1}\oplus W_{2}\text{,}%
\end{equation}
is a resonant subalgebra of $S_{E}^{\left(  4\right)  }\times\mathfrak{g}$,
where%
\begin{align}
W_{0}=\left(  S_{0}\times V_{0}\right)   &  =\left\{  \lambda_{0}\tilde
{J}_{ab},\lambda_{2}\tilde{J}_{ab},\lambda_{4}\tilde{J}_{ab},\lambda_{5}%
\tilde{J}_{ab}\right\}  ,\\
W_{1}=\left(  S_{1}\times V_{1}\right)   &  =\left\{  \lambda_{1}\tilde
{Q}_{\alpha},\lambda_{3}\tilde{Q}_{\alpha},\lambda_{5}\tilde{Q}_{\alpha
}\right\}  ,\\
W_{2}=\left(  S_{2}\times V_{2}\right)   &  =\left\{  \lambda_{2}\tilde{P}%
_{a},\lambda_{4}\tilde{P}_{a},\lambda_{5}\tilde{P}_{a}\right\}  .
\end{align}
Imposing the $0_{S}$-reduction condition,
\begin{equation}
\lambda_{5}T_{A}=0_{s},
\end{equation}
we find a new Lie superalgebra generated by $\left\{  J_{ab},P_{a},\tilde
{Z}_{ab},Z_{ab},\tilde{Z}_{a},Q_{\alpha},\Sigma_{\alpha}\right\}  $ where
these generators can be written as%
\begin{equation}%
\begin{array}
[c]{cc}%
J_{ab}=\lambda_{0}\tilde{J}_{ab}, & \tilde{Z}_{a}=\lambda_{2}\tilde{P}_{a},\\
\tilde{Z}_{ab}=\lambda_{2}\tilde{J}_{ab}, & Q_{\alpha}=\lambda_{1}\tilde
{Q}_{\alpha},\\
Z_{ab}=\lambda_{4}\tilde{J}_{ab}, & \Sigma_{\alpha}=\lambda_{3}\tilde
{Q}_{\alpha},\\
P_{a}=\lambda_{2}\tilde{P}_{a}, &
\end{array}
\end{equation}
and satisfy the following (anti)commutation relations%

\begin{align}
\left[  J_{ab},J_{cd}\right]   &  =\eta_{bc}J_{ad}-\eta_{ac}J_{bd}-\eta
_{bd}J_{ac}+\eta_{ad}J_{bc},\\
\left[  J_{ab},P_{c}\right]   &  =\eta_{bc}P_{a}-\eta_{ac}P_{b},\text{
\ \ \ \ \ \ \ }\left[  P_{a},P_{b}\right]  =Z_{ab},\\
\left[  J_{ab},Z_{cd}\right]   &  =\eta_{bc}Z_{ad}-\eta_{ac}Z_{bd}-\eta
_{bd}Z_{ac}+\eta_{ad}Z_{bc},\\
\left[  P_{a},Q_{\alpha}\right]   &  =-\frac{1}{2}\left(  \Gamma_{a}%
\Sigma\right)  _{\alpha},\\
\left[  J_{ab},Q_{\alpha}\right]   &  =-\frac{1}{2}\left(  \Gamma
_{ab}Q\right)  _{\alpha},\\
\left[  J_{ab},\Sigma_{\alpha}\right]   &  =-\frac{1}{2}\left(  \Gamma
_{ab}\Sigma\right)  _{\alpha},\\
\left\{  Q_{\alpha},Q_{\beta}\right\}   &  =-\frac{1}{2}\left[  \left(
\Gamma^{ab}C\right)  _{\alpha\beta}\tilde{Z}_{ab}-2\left(  \Gamma^{a}C\right)
_{\alpha\beta}P_{a}\right]  ,\\
\left\{  Q_{\alpha},\Sigma_{\beta}\right\}   &  =-\frac{1}{2}\left[  \left(
\Gamma^{ab}C\right)  _{\alpha\beta}Z_{ab}-2\left(  \Gamma^{a}C\right)
_{\alpha\beta}\tilde{Z}_{a}\right]
\end{align}%
\begin{align}
\left[  J_{ab},\tilde{Z}_{ab}\right]   &  =\eta_{bc}\tilde{Z}_{ad}-\eta
_{ac}\tilde{Z}_{bd}-\eta_{bd}\tilde{Z}_{ac}+\eta_{ad}\tilde{Z}_{bc},\\
\left[  \tilde{Z}_{ab},\tilde{Z}_{cd}\right]   &  =\eta_{bc}Z_{ad}-\eta
_{ac}Z_{bd}-\eta_{bd}Z_{ac}+\eta_{ad}Z_{bc},\\
\left[  J_{ab},\tilde{Z}_{c}\right]   &  =\eta_{bc}\tilde{Z}_{a}-\eta
_{ac}\tilde{Z}_{b},\text{ \ \ \ }\left[  \tilde{Z}_{ab},P_{c}\right]
=\eta_{bc}\tilde{Z}_{a}-\eta_{ac}\tilde{Z}_{b},\\
\left[  \tilde{Z}_{ab},Q_{\alpha}\right]   &  =-\frac{1}{2}\left(  \gamma
_{ab}\Sigma\right)  _{\alpha},\\
\text{others}  &  =0.
\end{align}
Here we have used the multiplication law of the semigroup $\left(
\ref{ls1}\right)  $ and the commutation relations of the original superalgebra
$\left(  \ref{ads1}-\ref{ads6}\right)  $. The new superalgebra obtained after
a $0_{S}$-reduced resonant $S$-expansion of $\mathfrak{osp}\left(  2|1\right)
\mathfrak{\otimes sp}\left(  2\right)  $ corresponds to the minimal Maxwell
superalgebra $s\mathcal{M}_{3}$. \ This superalgebra can be seen as the
supersymmetric extension of the generalized Maxwell algebra $g\mathcal{M}$ in
$D=3$ dimensions \cite{CR1}. \

Let us note that if we set $\tilde{Z}_{ab}=\tilde{Z}_{c}=0$, we obtain the
usual minimal Maxwell superalgebra \cite{BGKL}. \ As was pointed out in ref.
\cite{CR1}, this can be done since the Jacobi identities for the spinorial
generators are satisfied due to the gamma matrix identity $\left(  C\gamma
^{a}\right)  _{\left(  \alpha\beta\right.  }\left(  C\gamma_{a}\right)
_{\left.  \gamma\delta\right)  }=0$.

One can see that the minimal Maxwell superalgebra $s\mathcal{M}_{3}$ contains
the Maxwell algebra $\mathcal{M}=\left\{  J_{ab},P_{a},Z_{ab}\right\}  $ and
the Lorentz type algebra $\mathcal{L}^{\mathcal{M}}=\left\{  J_{ab}%
,Z_{ab}\right\}  $ introduced in ref. \cite{CPRS1} as subalgebras.

\subsection{Maxwell CS supergravity action in $D=3$}

\qquad Here, following the definitions and properties of the $S$-expansion
method, we present the construction of a three-dimensional Maxwell
supergravity action in the Chern-Simons formalism.

\qquad In order to write down an CS action for the minimal Maxwell
superalgebra $s\mathcal{M}_{3}$ we start from the one-form gauge connection%
\begin{align}
A  &  =\frac{1}{2}\omega^{ab}J_{ab}+\frac{1}{2}\tilde{k}^{ab}\tilde{Z}%
_{ab}+\frac{1}{2}k^{ab}Z_{ab}+\frac{1}{l}e^{a}P_{a}+\frac{1}{l}\tilde{h}%
^{a}\tilde{Z}_{a}+\frac{1}{\sqrt{l}}\psi^{\alpha}Q_{\alpha}+\frac{1}{\sqrt{l}%
}\xi^{\alpha}\Sigma_{\alpha},\nonumber\\
&  \label{connection}%
\end{align}
where the 1-form gauge fields are given in terms of the components of the
$\mathfrak{osp}\left(  2|1\right)  \otimes\mathfrak{sp}\left(  2\right)  $
connection $\tilde{e}^{a},\tilde{\omega}^{ab}$ and $\tilde{\psi}$:%
\[%
\begin{tabular}
[c]{llll}%
$\omega^{ab}=\lambda_{0}\tilde{\omega}^{ab},$ & $\tilde{k}^{ab}=\lambda
_{2}\tilde{\omega}^{ab}$ & $k^{ab}=\lambda_{4}\tilde{\omega}^{ab},$ &
$e^{a}=\lambda_{2}\tilde{e}^{a},$\\
$\tilde{h}^{a}=\lambda_{4}\tilde{e}^{a},$ & $\psi^{\alpha}=\lambda_{1}%
\tilde{\psi}^{\alpha},$ & $\xi^{\alpha}=\lambda_{3}\tilde{\psi}^{\alpha}.$ &
\end{tabular}
\]

The associated curvature two-form $F=dA+A\wedge A$ is given by
\begin{align}
F  &  =F^{A}T_{A}=\frac{1}{2}R^{ab}J_{ab}+\frac{1}{l}R^{a}P_{a}+\frac{1}%
{2}\tilde{F}^{ab}\tilde{Z}_{ab}+\frac{1}{2}F^{ab}Z_{ab}+\frac{1}{l}\tilde
{H}^{a}\tilde{Z}_{a}\nonumber\\
&  +\frac{1}{\sqrt{l}}\Psi^{\alpha}Q_{\alpha}+\frac{1}{\sqrt{l}}\Xi^{\alpha
}\Sigma_{\alpha}, \label{curvatura}%
\end{align}
where%
\begin{align*}
R^{ab}  &  =d\omega^{ab}+\omega_{\text{ }c}^{a}\omega^{cb},\\
R^{a}  &  =de^{a}+\omega_{\text{ }b}^{a}e^{b}-\frac{1}{2}\bar{\psi}\Gamma
^{a}\psi,\\
\tilde{F}^{ab}  &  =d\tilde{k}^{ab}+\omega_{\text{ }c}^{a}\tilde{k}%
^{cb}-\omega_{\text{ }c}^{b}\tilde{k}^{ca}+\frac{1}{2l}\bar{\psi}\Gamma
^{ab}\psi,\\
F^{ab}  &  =dk^{ab}+\omega_{\text{ }c}^{a}k^{cb}-\omega_{\text{ }c}^{b}%
k^{ca}+\tilde{k}_{\text{ }c}^{a}\tilde{k}^{cb}+\frac{1}{l^{2}}e^{a}e^{b}%
+\frac{1}{l}\bar{\xi}\Gamma^{ab}\psi,\\
\tilde{H}^{a}  &  =d\tilde{h}^{a}+\omega_{\text{ }b}^{a}\tilde{h}^{b}%
+\tilde{k}_{\text{ }c}^{a}e^{c}-\bar{\xi}\Gamma^{a}\psi,
\end{align*}%
\begin{align*}
\Psi &  =d\psi+\frac{1}{4}\omega_{ab}\Gamma^{ab}\psi,\\
\Xi &  =d\xi+\frac{1}{4}\omega_{ab}\Gamma^{ab}\xi+\frac{1}{4}\tilde{k}%
_{ab}\Gamma^{ab}\psi+\frac{1}{2l}e^{a}\Gamma_{a}\psi.
\end{align*}

The non-vanishing components of an invariant tensor for the Maxwell
superalgebra can be derived using the definitions of the $S$-expansion.
\ Indeed, by Theorem VII.2 of ref. \cite{Sexp}, \ the invariant tensor of an
$S$-expanded (super)algebra $\mathfrak{G}$ is given in terms of an invariant
tensor of the original (super)algebra $\mathfrak{g}$ through%
\begin{equation}
\left\langle T_{\left(  A,\alpha\right)  }T_{\left(  B,\beta\right)
}\right\rangle _{\mathfrak{G}}=\tilde{\alpha}_{\gamma}K_{\alpha\beta}^{\text{
\ \ }\gamma}\left\langle T_{A}T_{B}\right\rangle _{\mathfrak{g}},
\end{equation}
where $\tilde{\alpha}_{\gamma}$ are arbitrary constants and $K_{\alpha\beta
}^{\text{ \ \ }\gamma}$ corresponds to a $2$-selector. Thus, it is possible to
show that the only non-zero components of a symmetric invariant tensor for the
Maxwell superalgebra $s\mathcal{M}_{3}$ are given by%
\begin{align}
\left\langle J_{ab}J_{cd}\right\rangle _{s\mathcal{M}_{3}}  &  =\tilde{\alpha
}_{0}\left\langle \tilde{J}_{ab}\tilde{J}_{cd}\right\rangle =\alpha_{0}\left(
\eta_{ad}\eta_{bc}-\eta_{ac}\eta_{bd}\right)  ,\label{inv01}\\
\text{\ \ }\left\langle J_{ab}\tilde{Z}_{cd}\right\rangle _{s\mathcal{M}_{3}}
&  =\tilde{\alpha}_{2}\left\langle \tilde{J}_{ab}\tilde{J}_{cd}\right\rangle
=\alpha_{2}\left(  \eta_{ad}\eta_{bc}-\eta_{ac}\eta_{bd}\right)  ,\\
\left\langle \tilde{Z}_{ab}\tilde{Z}_{cd}\right\rangle _{s\mathcal{M}_{3}}  &
=\tilde{\alpha}_{4}\left\langle \tilde{J}_{ab}\tilde{J}_{cd}\right\rangle
=\alpha_{4}\left(  \eta_{ad}\eta_{bc}-\eta_{ac}\eta_{bd}\right)  ,\\
\left\langle J_{ab}Z_{cd}\right\rangle _{s\mathcal{M}_{3}}  &  =\tilde{\alpha
}_{4}\left\langle \tilde{J}_{ab}\tilde{J}_{cd}\right\rangle =\alpha_{4}\left(
\eta_{ad}\eta_{bc}-\eta_{ac}\eta_{bd}\right)  ,\\
\left\langle J_{ab}P_{c}\right\rangle _{s\mathcal{M}_{3}}  &  =\tilde{\alpha
}_{2}\left\langle \tilde{J}_{ab}\tilde{P}_{c}\right\rangle =\alpha_{1}%
\epsilon_{abc},\\
\left\langle \tilde{Z}_{ab}P_{c}\right\rangle _{s\mathcal{M}_{3}}  &
=\tilde{\alpha}_{4}\left\langle \tilde{J}_{ab}\tilde{P}_{c}\right\rangle
=\left\langle J_{ab}\tilde{Z}_{c}\right\rangle =\alpha_{3}\epsilon_{abc},\\
\left\langle P_{a}P_{b}\right\rangle _{s\mathcal{M}_{3}}  &  =\tilde{\alpha
}_{4}\left\langle \tilde{P}_{a}\tilde{P}_{b}\right\rangle =\alpha_{4}\eta
_{ab},\\
\left\langle Q_{\alpha}Q_{\beta}\right\rangle _{s\mathcal{M}_{3}}  &
=\tilde{\alpha}_{2}\left\langle \tilde{Q}_{\alpha}\tilde{Q}_{\beta
}\right\rangle =\left(  \alpha_{2}-\alpha_{1}\right)  C_{\alpha\beta},\\
\left\langle Q_{\alpha}\Sigma_{\beta}\right\rangle _{s\mathcal{M}_{3}}  &
=\tilde{\alpha}_{4}\left\langle \tilde{Q}_{\alpha}\tilde{Q}_{\beta
}\right\rangle =\left(  \alpha_{4}-\alpha_{3}\right)  C_{\alpha\beta}.
\label{inv08}%
\end{align}
where we have used eqs. $\left(  \ref{Inv1}\right)  $-$\left(  \ref{Inv4}%
\right)  $ and the following definitions%
\begin{align*}
\alpha_{0}  &  \equiv\tilde{\alpha}_{0}\mu_{0},\text{ \ \ }\alpha_{1}%
\equiv\tilde{\alpha}_{2}\mu_{1},\text{ \ \ \ }\alpha_{2}\equiv\tilde{\alpha
}_{2}\mu_{0}\text{\ }\\
\alpha_{3}  &  \equiv\tilde{\alpha}_{4}\mu_{1},~\ \ \alpha_{4}\equiv
\tilde{\alpha}_{4}\mu_{0}.
\end{align*}
Considering $\left(  \ref{inv01}-\ref{inv08}\right)  $ and the one-form
connection $\left(  \ref{connection}\right)  $in the general expression for
the Chern-Simons action $\left(  \ref{actm}\right)  $ we\ can write the CS
supergravity action for the minimal Maxwell superalgebra $s\mathcal{M}_{3}$
as
\begin{align}
S_{CS}^{\left(  2+1\right)  }  &  =k\int_{M}\left[  \frac{\mathbf{\alpha}_{0}%
}{2}\left(  \omega_{\text{ }b}^{a}d\omega_{\text{ }a}^{b}+\frac{2}{3}%
\omega_{\text{ }c}^{a}\omega_{\text{ }b}^{c}\omega_{\text{ }a}^{b}\right)
+\frac{\mathbf{\alpha}_{1}}{l}\left(  \epsilon_{abc}R^{ab}e^{c}-\bar{\psi}%
\Psi\right)  \right. \nonumber\\
&  +\mathbf{\alpha}_{2}\left(  R_{\text{ }b}^{a}\tilde{k}_{\text{ }a}%
^{b}+\frac{1}{l}\bar{\psi}\Psi\right)  +\frac{\mathbf{\alpha}_{3}}{l}\left(
\epsilon_{abc}\left(  R^{ab}\tilde{h}^{c}+D_{\omega}\tilde{k}^{ab}%
e^{c}\right)  -\bar{\xi}\Psi-\bar{\psi}\Xi\right) \nonumber\\
&  +\mathbf{\alpha}_{4}\left(  R_{\text{ }b}^{a}k_{\text{ }a}^{b}+\frac
{1}{l^{2}}e^{a}T_{a}+\frac{1}{l}\bar{\xi}\Psi+\frac{1}{l}\bar{\psi}\Xi\right)
\nonumber\\
&  \left.  -d\left(  \frac{\mathbf{\alpha}_{1}}{2l}\epsilon_{abc}\omega
^{ab}e^{c}+\frac{\mathbf{\alpha}_{3}}{2l}\epsilon_{abc}\left(  \tilde{k}%
^{ab}e^{c}+\omega^{ab}\tilde{h}^{c}\right)  +\frac{\mathbf{\alpha}_{2}}%
{2}\omega_{\text{ }b}^{a}\tilde{k}_{\text{ }a}^{b}+\frac{\mathbf{\alpha}_{4}%
}{2}\omega_{\text{ }b}^{a}k_{\text{ }a}^{b}\right)  \right]  . \label{ACT}%
\end{align}
where $T^{a}=D_{\omega}e^{a}$ is the torsion 2-form. \ The
Maxwell-Chern-Simons supergravity action $\left(  \ref{ACT}\right)  $ contains
four sectors proportional to different arbitrary constants $\alpha_{\gamma}$.
The first term corresponds to the so called exotic Lagrangian and it is
Lorentz invariant \cite{Witten, Zan}. \ The second term describes pure
supergravity without cosmological constant. \ On the other hand, the terms
proportional to $\alpha_{2}$, $\alpha_{3}$ and $\alpha_{4}$ contain the
coupling of the spin connection to the new gauge fields $\tilde{k}^{ab}$,
$k^{ab}$ and $\tilde{h}^{c}$. \ In particular the new Majorana spinor field
$\xi$ appears in the terms proportional to $\alpha_{3}$ and $\alpha_{4}$.
\ One can see that the bosonic part of the action $\left(  \ref{ACT}\right)  $
contains the CS gravity action found in ref. \cite{Topgrav, HRA}.

Let us note that the new fields appear also in the boundary term. \ Although
the boundary terms have no contribution to the dynamics of the theory, they
are an essential tool in the study of the $AdS/CFT$ correspondence
\cite{Maldacena, GKP, Witten2, AGMOO}. \ The inclusion of boundary
contributions to (super)gravity has been extensively studied in refs.
\cite{ACOTZ, MOTZ, NV, ADauria}.

Up to boundary terms, the full action $\left(  \ref{ACT}\right)  $ is
invariant under the local gauge transformations of the generalized Maxwell
supergroup,%
\begin{align}
\delta\omega^{ab}  &  =D_{\omega}\rho^{ab},\text{ \ \ \ \ \ }\delta
e^{a}=D_{\omega}\rho^{a}\,+e^{b}\rho_{b}^{\text{ }a}+\bar{\epsilon}\gamma
^{a}\psi,\label{ST01}\\
\delta\tilde{k}^{ab}  &  =D_{\omega}\tilde{\kappa}^{ab}-\left(  \tilde
{k}_{\text{ }c\text{ }}^{a}\rho_{\text{ }c}^{b}-\tilde{k}^{bc}\rho_{\text{ }%
c}^{a}\right)  -\frac{1}{l}\bar{\epsilon}\gamma^{ab}\psi,\\
\delta k^{ab}  &  =D_{\omega}\kappa^{ab}-\left(  k^{ac}\rho_{\text{ }c}%
^{b}-k^{bc}\rho_{\text{ }c}^{a}\right)  -\left(  \tilde{k}^{ac}\tilde{\kappa
}_{\text{ }c}^{b}-\tilde{k}^{bc}\tilde{\kappa}_{\text{ }c}^{a}\right)
\nonumber\\
&  +\frac{2}{l^{2}}e^{a}\rho^{b}-\frac{1}{l}\bar{\varrho}\gamma^{ab}\psi
-\frac{1}{l}\bar{\epsilon}\gamma^{ab}\xi,\\
\delta\tilde{h}^{a}  &  =D_{\omega}\tilde{\rho}^{a}+\tilde{h}^{b}\rho
_{b}^{\text{ }a}+\tilde{\kappa}_{\text{ }c}^{a}e^{c}+\tilde{k}_{\text{
}c\text{ }}^{a}\rho^{c}\,+\bar{\varrho}\gamma^{a}\psi+\bar{\epsilon}\gamma
^{a}\xi\\
\delta\psi &  =d\epsilon+\frac{1}{4}\omega^{ab}\gamma_{ab}\epsilon-\frac{1}%
{4}\rho^{ab}\gamma_{ab}\psi,\\
\delta\xi &  =d\varrho+\frac{1}{4}\omega^{ab}\gamma_{ab}\varrho+\frac{1}%
{2l}e^{a}\gamma_{a}\epsilon-\frac{1}{2l}\rho^{a}\gamma_{a}\psi-\frac{1}{4}%
\rho^{ab}\gamma_{ab}\xi\nonumber\label{ST06}\\
&  +\frac{1}{4}\tilde{k}^{ab}\gamma_{ab}\epsilon-\frac{1}{4}\tilde{\kappa
}^{ab}\gamma_{ab}\psi.
\end{align}
where the $s\mathcal{M}_{3}$ gauge parameter is given by%
\begin{equation}
\rho=\frac{1}{2}\rho^{ab}J_{ab}+\frac{1}{2}\tilde{\kappa}^{ab}\tilde{Z}%
_{ab}+\frac{1}{2}\kappa^{ab}Z_{ab}+\frac{1}{l}\rho^{a}P_{a}+\frac{1}{l}%
\tilde{\rho}^{a}\tilde{Z}_{a}+\frac{1}{\sqrt{l}}\epsilon^{\alpha}Q_{\alpha
}+\frac{1}{\sqrt{l}}\varrho^{\alpha}\Sigma_{\alpha}.
\end{equation}

\section{Comments and possible developments}

\qquad In the present work we have derived the $D=3$ $\mathcal{N}=1$
Chern-Simons supergravity action from the minimal Maxwell superalgebra
$s\mathcal{M}_{3}$. \ We have shown that the Maxwell supersymmetries can be
obtained from the $\mathfrak{osp}\left(  2|1\right)  \otimes\mathfrak{sp}%
\left(  2\right)  $ superalgebra using the semigroup expansion procedure. The
method considered here allowed to obtain the invariant tensor for the Maxwell
superalgebra and to build the most general $D=3$ CS supergravity action
invariant under the Maxwell supergroup. \ The action describes a supergravity
theory without cosmological constant in three dimensions and can be seen as a
supersymmetric extension of the results in refs. \cite{Topgrav, HRA} where new
extra fields have been added in order to have well defined $S$-expanded
invariant tensors.

It is interesting to note that the minimal Maxwell superalgebra can also be
derived as a generalized IW contraction of a minimal $AdS$-Lorentz
superalgebra \cite{CRS}. \ Analogously, the non-standard Maxwell superalgebra
introduced in refs. \cite{Sorokas, GIW, Sorokas3}, can be recovered by
performing a suitable IW contraction of the usual supersymmetric extension of
the $AdS$-Lorentz algebra
\footnote{Also known as Poincar\'{e}
semi-simple extended or $\mathfrak{so}(D-1,1)\oplus\mathfrak{so}%
(D-1,2)$ algebra.
The semi-simple $o(N)$-extended superPoincar\'{e}
algebra can be found in ref. \cite{Sorokas2}.}
\cite{Sorokas, DKGS}. \ Additionally, as shown in refs. \cite{DFIMRSV, SS,
FISV}, the $AdS$-Lorentz (super)algebra can be alternatively obtained as an
$S$-expansion of the $AdS$ (super)algebra . \ Then, one could construct a CS
supersymmetric action from the non-standard Maxwell superalgebra combining the
$S$-expansion method with the IW contraction. \ Nevertheless,\ in this
superalgebra the four-momentum generators $P_{a}$ are not expressed as
bilinears expressions of fermionic generators $Q$ $\left(  \left\{  Q_{\alpha
},Q_{\beta}\right\}  =-\frac{1}{2}\left(  \Gamma^{ab}C\right)  _{\alpha\beta
}Z_{ab}\right)  .$ \ As a consequence, the supersymmetric action constructed
out of the non-standard Maxwell superalgebra, shall not describe a
supergravity action but an exotic alternative supersymmetric action.

Our results provide one more example of the usefulness of the Maxwell
(super)symmetry in (super)gravity (see \cite{CPRS1, CPRS2, CR2, Topgrav,
HRA}). \ In particular, we have shown that the semigroup expansion procedure
can be used in order to derive a new supergravity theory. \ The same procedure
could be used in eleven dimensions in order to recover the CJS supergravity
theory from a given superalgebra.

It would be interesting to analyze the boundary contributions present in the
CS supergravity action in the $AdS/CFT$ context. \ On the other hand, the
results presented here could be useful in the construction of supergravity
actions in higher dimensions. \ It seems that it should be possible to recover
standard odd-dimensional supergravity from the Maxwell supersymmetries [work
in progress].

A future work could be consider the $\mathcal{N}$-extended Maxwell
superalgebras and their generalizations in order to build $\left(  p,q\right)
$-type CS supergravity models in a very similar way to the one introduced
here. \ Eventually, the semigroup expansion could be useful in the
construction of matter-supergravity theories.

\section*{Acknowledgements \textbf{\ }}

This work was supported in part by FONDECYT Grants N$%
{{}^\circ}%
$ 1130653. Two of the authors (P.K. C., E.K. R.) were supported by grants from
the Comisi\'{o}n Nacional de Investigaci\'{o}n Cient\'{\i}fica y
Tecnol\'{o}gica (CONICYT) and from the Universidad de Concepci\'{o}n, Chile.
\ P.K. C. and E.K. R. wish to thank L. Andrianopoli, R. D'Auria and M.
Trigiante for their kind hospitality at Dipartimieto Scienza Applicata e
Tecnologia (DISAT) of Politecnico di Torino.


\begin{thebibliography}{99}                                                                                               %


\bibitem {BCR}H. Bacry, P. Combe, J.L. Richard, \textit{Group-theoretical
analysis of elementary particles in an external electromagnetic field. 1. The
relativistic particle in a constant and uniform field}, Nuovo Cim. A
\textbf{67} (1970) 267.

\bibitem {Schrader}R. Schrader, \textit{The Maxwell group and the quantum
theory of particles in classical homogeneous electromagnetic fields}, Fortsch.
Phys. \textbf{20} (1972) 701.

\bibitem {AKL}J.A. de Azcarraga, K. Kamimura, J. Lukierski,
\textit{Generalized cosmological term from Maxwell symmetries}, Phys. Rev. D
\textbf{83} (2011) 124036. arXiv:1012.4402 [hep-th].

\bibitem {GRCS}F. Izaurieta, P. Minning, A. Perez, E. Rodr\'{\i}guez, P.
Salgado, \textit{Standard General Relativity from Chern-Simons Gravity}, P.
Salgado,\textit{\ }Phys. Lett. B \textbf{678,} 213\ (2009). arXiv:0905.2187 [hep-th].

\bibitem {CPRS1}P.K. Concha, D.M. Pe\~{n}afiel, E.K. Rodr\'{\i}guez, P.
Salgado, \textit{Even-dimensional General Relativity from Born-Infeld
gravity}, Phys. Lett. B \textbf{725}, 419 (2013). arXiv:1309.0062 [hep-th].

\bibitem {CPRS2}P.K. Concha, D.M. Pe\~{n}afiel, E.K. Rodr\'{\i}guez, P.
Salgado, \textit{Chern-Simons and Born-Infeld gravity theories and Maxwell
algebras type}, Eur. Phys. J. C \textbf{74} (2014) 2741. arXiv:1402.0023 [hep-th].

\bibitem {CPRS3}P.K. Concha, D.M. Pe\~{n}afiel, E.K. Rodr\'{\i}guez, P.
Salgado, \textit{Generalized Poincare algebras and Lovelock-Cartan gravity
theory}, Phys. Lett. B \textbf{742} (2015) 310. arXiv:1405.7078 [hep-th].

\bibitem {BGKL}S. Bonanos, J. Gomis, K. Kamimura and J. Lukierski,
\textit{Maxwell Superalgebra and Superparticle in Constant Gauge Backgrounds},
Phys. Rev. Lett. \textbf{104} (2010) 090401. arXiv:0911.5072 [hep-th].

\bibitem {AILW}J.A. de Azcarraga, J.M. Izquierdo, J. Lukierski, M. Woronowicz,
\textit{Generalizations of Maxwell (super)algebras by the expansion method},
Nucl. Phys. B \textbf{869} (2013) 303. arXiv:1210.1117 [hep-th].

\bibitem {CR1}P.K. Concha, E.K. Rodr\'{\i}guez, \textit{Maxwell Superalgebras
and Abelian Semigroup Expansion}, Nucl. Phys. B \textbf{886} (2014) 1128.
arXiv:1405.1334 [hep-th].

\bibitem {AF}R. D'Auria, P. Fr\'{e}, \textit{Geometric Supergravity in d=11
and Its Hidden Supergroup}, Nucl. Phys. B \textbf{201} (1982) 101.

\bibitem {Green}M.B. Green, \textit{Supertranslations, Superstrings and
Chern-Simons Forms}, Phys. Lett. B \textbf{223} (1989) 157.

\bibitem {CR2}P.K. Concha, E.K. Rodr\'{\i}guez, \textit{N=1 supergravity and
Maxwell superalgebras}, JHEP \textbf{1409} (2014) 090. arXiv:1407.4635 [hep-th].

\bibitem {Sexp}F. Izaurieta, E. Rodr\'{\i}guez, P. Salgado, \textit{Expanding
Lie (super)algebras through Abelian semigroups}, J. Math. Phys. \textbf{47}
(2006) 123512 [hep-th/0606215].

\bibitem {IRS1}F. Izaurieta, E. Rodr\'{\i}guez, P. Salgado,
\textit{Eleven-dimensional gauge theory for the M algebra as an Abelian
semigroup expansion of osp(32%
$\vert$%
1)}, Eur. Phys. J. C \textbf{54} (2008) 675 [arXiv:hep-th/0606225].

\bibitem {GSRS}N. Gonz\'{a}lez, P. Salgado, G. Rubio, S. Salgado,
\textit{Einstein-Hilbert action with cosmological term from Chern-Simons
gravity}, J. Geom. Phys. \textbf{86} (2014) 339.

\bibitem {Topgrav}P. Salgado, R. J. Szabo, O. Valdivia, \textit{Topological
gravity and transgression holography}, Phys. Rev. D \textbf{89} (2014) 084077.
arXiv:1401.3653 [hep-th].

\bibitem {AchuTown}A. Achucarro, P.K. Townsend, \textit{A Chern-Simons Action
for Three-Dimensional anti-De Sitter Supergravity Theories}, Phys. Lett. B
\textbf{180} (1986) 89.

\bibitem {Cham1}A.H. Chamseddine, \textit{Topological Gauge Theory of Gravity
in Five-dimensions and All Odd Dimensions}, Phys. Lett. B \textbf{223} (1989) 291.

\bibitem {Cham2}A.H. Chamseddine, \textit{Topological gravity and supergravity
in various dimensions}, Nucl. Phys. B \textbf{346} (1990) 213.

\bibitem {GTW}A. Giacomini, R. Troncoso, S. Willison,
\textit{Three-dimensional supergravity reloaded}, Class. Quant. Grav.
\textbf{24} (2007) 2845 \ [hep-th/0610077].

\bibitem {AMNT}L. Andrianopoli, N. Merino, F. Nadal, M. Trigiante,
\textit{General properties of the expansion methods of Lie algebras}, J. Phys.
A \textbf{46} (2013) 365204. arXiv:1308.4832 [gr-qc].

\bibitem {Witten}E. Witten, \textit{(2+1)-Dimensional gravity as an exactly
soluble system}, Nucl. Phys. B \textbf{311}, 46 (1988).

\bibitem {Zan}J. Zanelli, \textit{Lectures on Chern-Simons (super)gravities},
Second edition. arXiv:hep-th/0502193.

\bibitem {HRA}S. Hoseinzadeh, A. Rezaei-Aghdam, \textit{(2+1)-dimensional
gravity from Maxwell and semisimple extension of the Poincar\'{e} gauge
symmetric models}, Phys. Rev. D \textbf{90} (2014) 084008. arXiv:1402.0320 [hep-th].

\bibitem {Maldacena}J.M. Maldacena, \textit{The Large N limit of
superconformal field theories and supergravity}, Adv. Theor. Math. Phys.
\textbf{2} (1998) 231 [hep-th/9711200].

\bibitem {GKP}S.S. Gubser, I.R. Klebanov, A.M. Polyakov, \textit{Gauge theory
correlators from non-critical string theory}, Phys. Lett. B \textbf{428}
(1998) 105 [hep-th/9802109].

\bibitem {Witten2}E. Witten, \textit{Anti-de Sitter space and holography},
Adv. Theor. Math. Phys. \textbf{2} (1998) 253 [hep-th/9802150].

\bibitem {AGMOO}O. Ahorony, S.S. Gubser, J.M. Maldacena, H. Ooguri, Y. Oz,
\textit{Large N field theories, string theory and gravity}, Phys. Rept.
\textbf{323} (2000) 183 [hep-th/9905111].

\bibitem {ACOTZ}R. Aros, M. Contreras, R. Olea, R. Troncoso, J. Zanelli,
\textit{Conserved charges for gravity with locally AdS asymptotics}, Phys.
Rev. Lett. \textbf{84} (2000) 1647 [gr-qc/9909015].

\bibitem {MOTZ}P. Mora, R. Olea, R. Troncoso, J. Zanelli, \textit{Finite
action principle for Chern-Simons AdS gravity}, JHEP \textbf{0406} (2004) 036 [hep-th/0405267].

\bibitem {NV}P. van Nieuwenhuizen, D.V. Vassilevich, \textit{Consistent
boundary condition for supergravity}, Class. Quant. Grav. \textbf{22} (2005)
5029 [hep-th/0507172].

\bibitem {ADauria}L. Andrianopoli, R. D'Auria, \textit{N=1 and N=2 pure
supergravities on a manifold with boundary}, JHEP \textbf{1408} (2014) 012.
arXiv:1405.2010 [hep-th].

\bibitem {CRS}P.K. Concha, E.K. Rodr\'{\i}guez, P. Salgado,
\textit{Generalized supersymmetric cosmological term in N=1 Supergravity},
JHEP \textbf{1508} (2015) 009. arXiv:1504.01898 [hep-th].

\bibitem {Sorokas}D.V. Soroka, V.A. Soroka, \textit{Semi-simple extension of
the (super)Poincare algebra}, Adv. High Energy Phys. 2009 (2009) 34147 [hep-th/0605251].

\bibitem {GIW}J. Lukierski, \textit{Generalized Wigner-Inonu Contractions and
Maxwell (Super)Algebras}, Proc. Steklov Inst. Math. 272 (2011) 183.
arXiv:1007.3405 [hep-th].

\bibitem {Sorokas3}D.V. Soroka, V.A. Soroka, \textit{Tensor extension of the
Poincar\'{e} algebra}, Phys. Lett. B \textbf{607} (2005) 302 arXiv:hep-th/0410012.

\bibitem {Sorokas2}D.V. Soroka, V.A. Soroka, \textit{Semi-simple o(N)-extended
super-Poincar\'{e} algebra}, arXiv:1004.3194 [hep-th].

\bibitem {DKGS}R. Durka, J. Kowalski-Glikman, M. Szczachor, \textit{Gauged
AdS-Maxwell algebra and gravity}, Mod. Phys. Lett. A \textbf{26} (2011) 2689.
arXiv:1107.4728 [hep-th].

\bibitem {DFIMRSV}J. Diaz, O. Fierro, F. Izaurieta, N. Merino, E.
Rodr\'{\i}guez, P. Salgado, O. Valdivia, \textit{A generalized action for (2 +
1)-dimensional Chern-Simons gravity}, J. Phys. A \textbf{45} (2012) 255207.
arXiv:1311.2215 [gr-qc].

\bibitem {SS}P. Salgado, S. Salgado, $\mathfrak{so}\left(  D-1,1\right)
\otimes\mathfrak{so}\left(  D-1,2\right)  $\textit{ algebras and gravity},
Phys. Lett. B \textbf{728}, 5 (2013).

\bibitem {FISV}O. Fierro, F. Izaurieta, P. Salgado, O. Valdivia,
\textit{(2+1)-dimensional supergravity invariant under the AdS-Lorentz
superalgebra}, arXiv:1401.3697 [hep-th].
\end{thebibliography}
\end{document}